\documentclass[sigconf,screen,authorversion]{acmart} 

\usepackage[utf8]{inputenc} 
\usepackage{csquotes} 
\usepackage{cleveref} 
\usepackage{multirow} 
\usepackage{enumitem}
\usepackage{balance}

\copyrightyear{2020}
\acmYear{2020}
\setcopyright{acmlicensed}
\acmConference[SIGCSE '20]{The 51st ACM Technical Symposium on Computer Science Education}{March 11--14, 2020}{Portland, OR, USA}
\acmBooktitle{The 51st ACM Technical Symposium on Computer Science Education (SIGCSE '20), March 11--14, 2020, Portland, OR, USA}
\acmPrice{15.00}
\acmDOI{10.1145/3328778.3366908}
\acmISBN{978-1-4503-6793-6/20/03}

\begin{document}

\title[KYPO4INDUSTRY: A Testbed for Teaching Cybersecurity of Industrial Control Systems]{KYPO4INDUSTRY: A Testbed for Teaching Cybersecurity \\ of Industrial Control Systems}

\author{Pavel Čeleda}
\affiliation{
  \institution{Masaryk University}
  \country{Czech Republic}
}
\email{celeda@ics.muni.cz}

\author{Jan Vykopal}
\orcid{0000-0002-3425-0951}
\affiliation{
  \institution{Masaryk University}
  \country{Czech Republic}
}
\email{vykopal@ics.muni.cz}

\author{Valdemar Švábenský}
\orcid{0000-0001-8546-280X}
\affiliation{
  \institution{Masaryk University}
  \country{Czech Republic}
}
\email{svabensky@ics.muni.cz}

\author{Karel Slavíček}
\affiliation{
  \institution{Masaryk University}
  \country{Czech Republic}
}
\email{slavicek@ics.muni.cz}

\newcommand{\KI}{{KYPO4INDUSTRY }}
\newcommand{\unipi}{{UniPi }}

\begin{abstract}
There are different requirements on cybersecurity of industrial control systems and information technology systems. This fact exacerbates the global issue of hiring cybersecurity employees with relevant skills. In this paper, we present \KI training facility and a course syllabus for beginner and intermediate computer science students to learn cybersecurity in a simulated industrial environment. The training facility is built using open-source hardware and software and provides reconfigurable modules of industrial control systems. The course uses a flipped classroom format with hands-on projects: the students create educational games that replicate real cyber attacks. Throughout the semester, they learn to understand the risks and gain capabilities to respond to cyber attacks that target industrial control systems. Our described experience from the design of the testbed and its usage can help any educator interested in teaching cybersecurity of cyber-physical systems.
\end{abstract}



\maketitle

\section{Introduction}



\begin{figure}[h]
    \centering
    \includegraphics[width=0.85\linewidth]{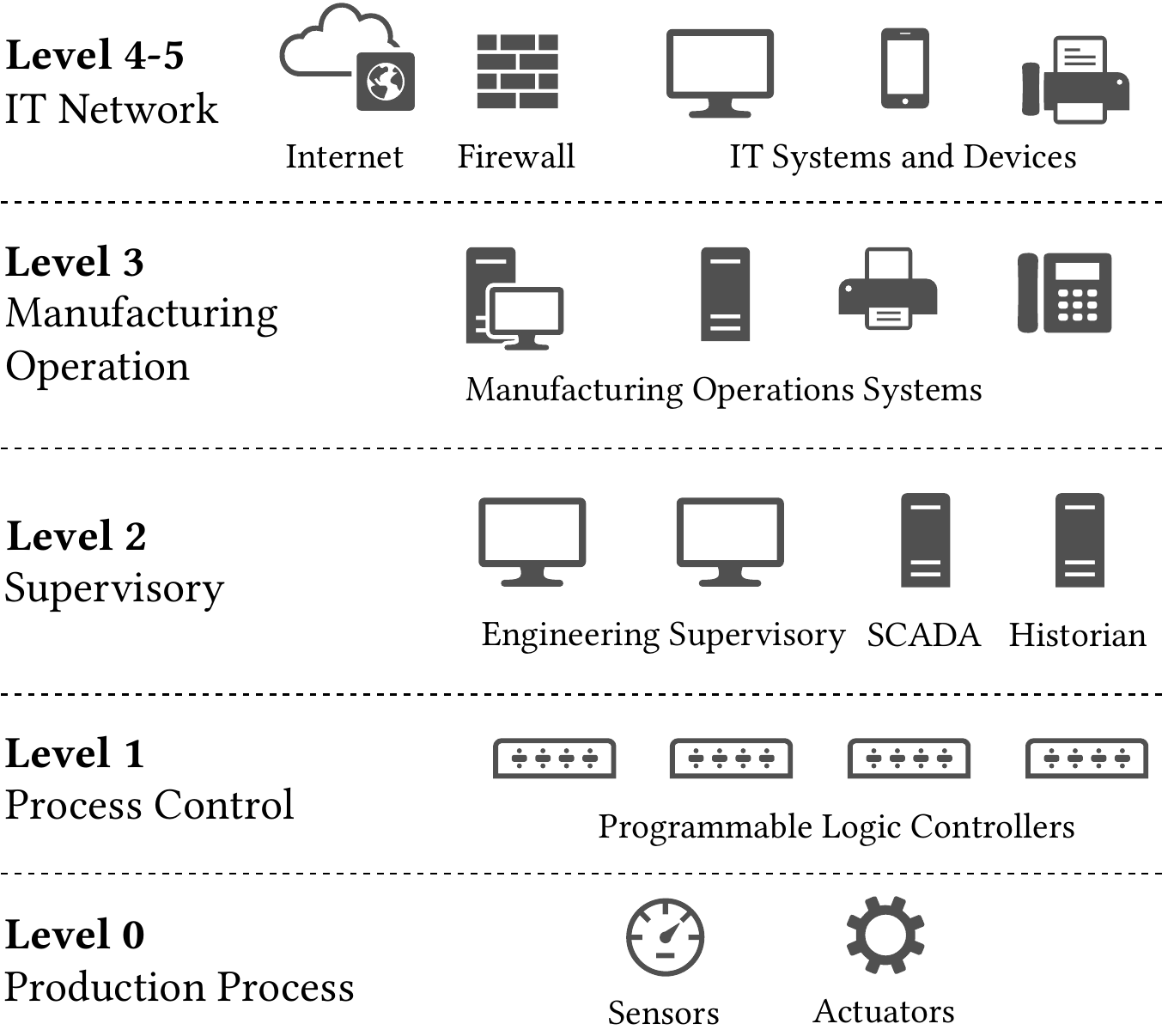}
    \caption{The ISA-95 architecture: A hierarchical model of enterprise-control system integration~\cite{scholten2007road}}
    \label{fig:isa-95-architecture}
\end{figure}

Industrial control systems (ICS) provide vital services, such as electricity, water treatment, and transportation. Although these systems were formerly isolated, they became connected with information technology (IT) systems and even to the Internet. \Cref{fig:isa-95-architecture} shows the ISA-95 enterprise reference architecture that describes the connection between the functions of ICS and IT systems~\cite{scholten2007road}. This connection of processes in the cyberspace and the physical world has reduced costs and enabled new services. However, the ICS assets became vulnerable to new threats and ever-evolving cyber threat landscape~\cite{zhu2011taxonomy}.

ICSs are made to maintain the integrity and availability of production processes and to sustain conditions of industrial environments. Their hardware and software components are often custom-built and tightly integrated. However, IT systems use off-the-shelf hardware and software and have different operational characteristics and security objectives~\cite{neitzel2014top}.

Traditional cybersecurity courses are falling short in training ICS security~\cite{butts2015industrial}, since they focus on exploiting and defending IT assets. To teach ICS security, a training facility (testbed) is needed to model a real-world ICS system~\cite{holm2015ICStestbedsurvey} and to provide hands-on experience. However, building and operating a realistic cyber-physical testbed using standard industrial equipment is expensive. It incorporates equipment such as programmable logic controllers (PLC), input/output modules, sensors, actuators, and other devices.


This paper addresses how to teach ICS cybersecurity to computer science students. Currently, the majority of students has intermediate knowledge level of IT cybersecurity but is unfamiliar with ICS principles. Our work brings two main contributions. First, we share our experience with the design and acquisition of \KI testbed. Second, we describe a course syllabus to deliver cybersecurity training in a simulated industrial environment. The course uses a flipped classroom format~\cite{bishop2013flipped} with hands-on projects replicating real cyber attacks. The students learn to understand the risks and gain capabilities to respond to cyber attacks that target ICS.


This paper is organized into five sections. \Cref{sec:related-work} provides an overview of hands-on activities for teaching cybersecurity in IT and ICS. \Cref{sec:testbed} describes the ICS training facility, lists the main components, and provides implementation details. \Cref{sec:ics-course} provides a detailed description of the design, content, and assessment methods of the ICS cybersecurity course. Finally, \Cref{sec:conclusions} concludes the paper and outlines future work.

\section{Related Work} \label{sec:related-work}


Cybersecurity knowledge and skills are usually taught through classroom lectures complemented with labs, exercises, and home assignments. Such a combination of theory and practice is essential in training cybersecurity experts, since the number of cyber attacks and the ingenuity of attackers is ever-growing. This section presents the current best practice for teaching cybersecurity in IT and ICS.

\subsection{Teaching Cybersecurity in IT} \label{sec:learning_by_doing} 

The three most popular types of IT cybersecurity training are hands-on assignments, capture the flag (CTF) games, and cyber defense exercises (CDX). Hands-on assignments include working with cybersecurity tools, usually in a virtual environment. An example collection of such assignments is SecKnitKit~\cite{siraj2015}, a set of virtual machines (VMs) and corresponding learning materials. Using ready-made VMs offers a realistic and isolated environment with minimal setup, which is well-suited for cybersecurity training. Alternatively, online learning platforms, such as Root Me~\cite{rootme}, provide a set of cybersecurity challenges that the learners solve locally or online.

CTF is a format of cybersecurity games and competitions in which the learners solve various cybersecurity tasks. Completing each task yields a textual string called flag, which is worth a certain amount of points. There are two main variations of the CTF format: Jeopardy and Attack-Defense.

In Jeopardy CTF, such as PicoCTF~\cite{chapman2014}, learners choose the tasks to solve from a static collection of challenges presented in a web interface. The challenges are divided into categories such as cryptography, reverse engineering, or forensics. Learners solve the tasks locally at their machines or interact with a remote server. Jeopardy CTFs can thus accommodate hundreds of players at the same time.

In Attack-Defense CTF, such as iCTF~\cite{vigna2014}, teams of learners each maintain an identical instance of a vulnerable computer network. Each team must protect its network while exploiting vulnerabilities in the networks of other teams. Successful attacks yield flags, which, along with maintaining the availability of the network services, contribute to the teams' score.

While anyone can participate in hands-on training or CTF games, CDX is a complex cybersecurity exercise for professionals, often from military or government agencies or dedicated cybersecurity teams~\cite{eagle2013,kypo-cdx}. Learners are divided into blue teams responsible for maintaining and defending a complex network infrastructure against attacks of an external red team. The blue teams must preserve the availability of the network services for end-users and respond to prompts from law enforcement groups and journalists. Beyond IT systems, some exercises feature simulated critical infrastructure (e.g., electricity grid or transportation).

\subsection{Teaching Cybersecurity in ICS}


Teaching ICS relies on components that are likely to be encountered in operational environments. Testbeds are built to replicate the behavior of ICS and incorporate a control center, communication architecture, field devices, and physical processes~\cite{stouffer2015nist}. Holm et al. surveyed the current ICS testbeds and reported on their objectives and implementation~\cite{holm2015ICStestbedsurvey}. Most testbeds focus on cybersecurity -- vulnerability analysis, tests of defense mechanisms, and education. Testbed fidelity is essential for training activities and the level of provided courses. High-fidelity testbeds are rare, and most testbeds use simulations, scaled-down models, and individual components~\cite{butts2015industrial}. ICS courses cover beginner and intermediate levels of training.

Virtualized, purely software-based testbeds are built upon virtual PLCs and devices modeled in software~\cite{alves2018virtscada}. They can be highly flexible and imitate any real environment with an arbitrary number of various devices. Their main drawback is the lack of look and feel of the operational environment. Users who are accustomed to using the real equipment might perceive purely software-based testbeds as a computer game and not as training for real situations. An example of such testbed is a system for assessment of cyber threats against networked critical infrastructures~\cite{Siaterlis:2014:CT:2602695.2602575}. 


Hardware-based testbeds are used, for example, in training operating personnel of chemical and nuclear plants. Apart from these, there are other specialized ones, such as PowerCyber~\cite{Hahn2010PowerCyber}, which is designed to closely resemble power grid communication utilizing actual field devices and Supervisory Control and Data Acquisition (SCADA) software. This testbed allows to explore cyber attacks and defenses while evaluating their impact on power flow. Ahmed et al.~\cite{Ahmed:2016:SST:3018981.3018984} presented SCADA testbed that demonstrates three industrial processes (a gas pipeline, a power transmission and distribution system, and a wastewater treatment plant) in a small scale. To do so, it employs real-world industrial equipment, such as PLCs, sensors, or aerators. These are deployed at each physical process system for local control and monitoring, and the PLCs are connected to a computer running human-machine interface (HMI) software for monitoring the status of the physical processes. The testbed is used in a university course on ICS security. Students can observe the industrial processes, learn ladder logic programming in various programming environments, and observe network traffic of multiple communication protocols.


In 2016, Antonioli et al.~\cite{Antonioli:2017:GIS:3140241.3140253} prepared SWaT Security Showdown, the first CTF event targeted at ICS security. The game employed Secure Water Treatment (SWaT), a software-based testbed available at Singapore University of Technology and Design~\cite{7469060}. Selected twelve international teams from academia and industry were invited. The game was divided into two phases: online Jeopardy and on-site Attack-Defense CTF. The first part served as a training session and included novel categories related to the ICS realm. The on-site CTF lasted two days. The teams visited the testbed on the first day. The next day, they had three hours to attack the SWaT testbed. The authors devised a dedicated scoring system for the assessment of attacks launched by the teams. The scoring evaluated the impact of the attacks on the physical and monitoring processes of the testbed, and the ability to conduct attacks that are not discovered by ICS detection systems deployed in the testbed.

Chothia and de Ruiter~\cite{198081} developed a course at the University of Birmingham on penetration testing techniques of off-the-shelf consumer Internet of Things (IoT) devices. Students were tasked to analyze device functionality, write up a report, and give a presentation of their findings. 

\section{KYPO4INDUSTRY: ICS Training Facility} 
\label{sec:testbed}

\begin{figure}[!ht]
    \centering
    \includegraphics[width=0.8\linewidth]{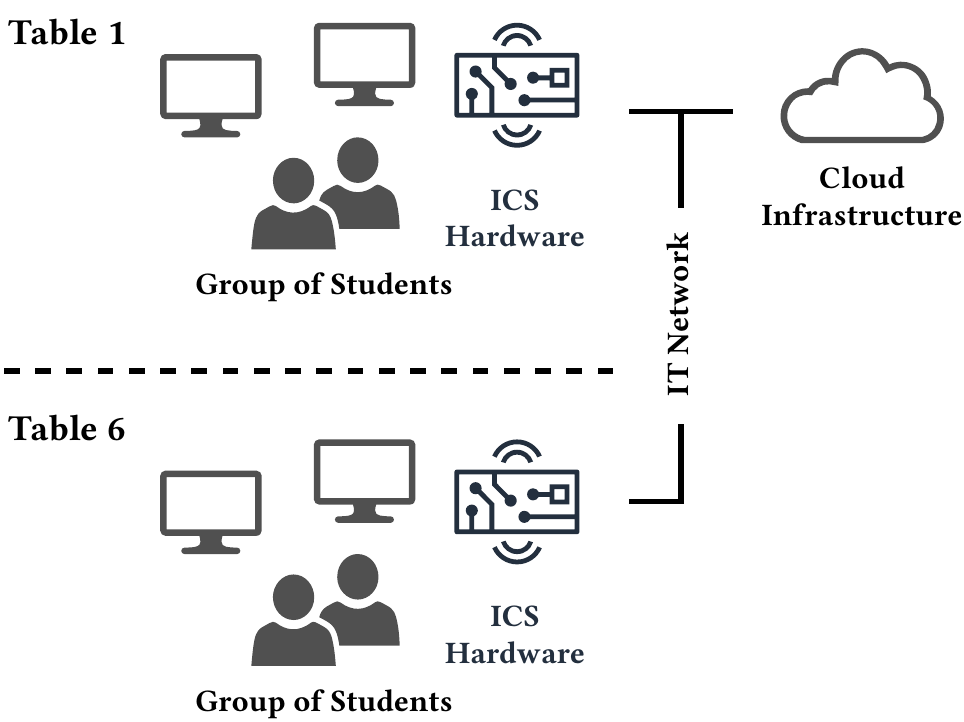}
    \caption{Training facility setup}
    \label{fig:training-facility}
\end{figure}

In this section, we describe the hardware and software components of the ICS testbed. The ICS training takes place in a specialized physical facility, which has been frequently used for university courses~\cite{kypolab-course}, international CDXs~\cite{kypo-cdx}, and extracurricular events. The room contains six large tables, each with three seats, three desktop PCs, and ICS hardware devices. As \Cref{fig:training-facility} shows, the devices within the testbed infrastructure are interconnected and so can communicate with each other. The tables are portable to allow the instructor to rearrange the room for various activities, including team assignments, student presentations, and group discussions.

\subsection{Hardware Components}


Based on the discussions with our partners and our experience, we defined these requirements on the hardware components of the \KI testbed:

\begin{itemize}[leftmargin=15pt]
\item \emph{Open-hardware} – full access to hardware and software to avoid vendor-lock and other proprietary limitations, unlimited software manipulation, and community support. 
\item \emph{Performance} – the PLC processor and memory (RAM, FLASH) must be sufficient to host operating-system with virtualization support (containers) and TCP/IP networking.
\item \emph{Communication interfaces} – wired and wireless communication buses for connecting peripherals and devices in the testbed. Industry standards like Ethernet, Wi-Fi, Bluetooth, USB, RS-485, and 1-Wire must cover both IT and ICS environments. 
\item \emph{Inputs} – digital inputs to read binary sensors and devices such as buttons, switches, and motion sensors. Analog inputs to measure voltage from temperature, pressure, and light sensors.
\item \emph{Outputs} – digital outputs to switch binary actuators (LEDs, relays, motors), seven-segment displays, and graphical display (touchscreen) for human-machine interface.
\item \emph{Physical dimension} – hardware setup which will provide a cyber-physical experience (allow manipulation and observation of physical processes), multiple devices mounted in the same control panel, tabletop and mobile setup.
\item \emph{Safety} – durable equipment and a tamper-resistant installation, all cabling and connectors should be concealed to prevent (un)intended tampering during hands-on training, and electrical safety – avoid grid power parts.
\end{itemize}

\begin{figure}[!ht]
    \centering
    \includegraphics[width=0.9\linewidth]{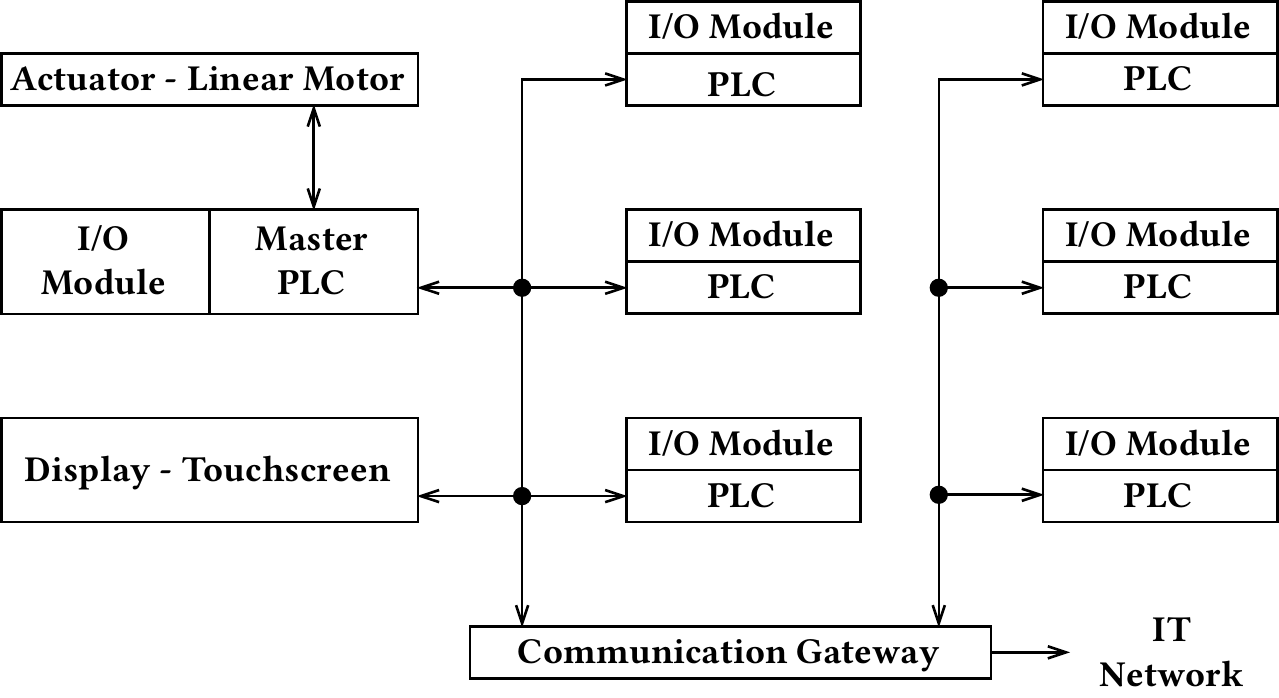}
    \caption{Control panel block diagram}
    \label{fig:testbed-hw-architecture}
\end{figure}

\Cref{fig:testbed-hw-architecture} shows the proposed hardware architecture. The hardware components of the control panel include PLCs, I/O modules, touchscreen, linear motor, and communication gateway.

PLC devices are a fundamental component of the control panel. When choosing a suitable PLC platform, it was essential for us that it leverages well-known hardware and has an industrial appearance. We chose the \unipi platform, which uses the popular Raspberry Pi single-board computer~\cite{raspberry} and industrial casing. The \unipi Neuron M103~\cite{unipiM103} model is used as the master PLC, and slave PLCs use \unipi Neuron S103~\cite{unipiS103}. Both versions use Raspberry Pi 3 Model B with four-core 1.2\,GHz CPU and 1\,GB RAM. The Neuron PLC is DIN rail mountable, requires 24\,V DC power supply, and has the following interfaces:

\begin{itemize}[leftmargin=15pt]
    \item{10/100 Mbit Ethernet, Wi-Fi, Bluetooth,}
    \item{four USB 2.0 ports, Micro SD port,}
    \item{RS-485, 1-Wire interface,}
    \item{digital input and output pins,}
    \item{one analog input and one analog output port.}
\end{itemize}


The control panel uses two I/O module types. The first one connects the master PLC to three large-area LEDs, two buttons, one key switch, and two motion detectors. The master PLC controls the linear motor through the RS-485 interface. The second type connects slave PLC with three large-area LEDs, two buttons, high power led (heating), 1-wire digital thermometer, and light sensor (analog input). Slave PLC uses RS-485 to control two-digit seven-segment display and 1.54" e-paper module.

10" LCD touchscreen is used to display technology processes. Dedicated Raspberry Pi module controls LCD via the HDMI interface and the touch panel via USB. A mechanical demonstrator (actuator) uses a linear motor. It includes DRV8825 stepper motor driver, ATmega 328 MCU, two end-stops switches, and three infrared position sensors. A network switch (MikroTik CRS125-24G) connects all PLC devices. Switch manages the flow of data between PLCs (100 Mbit Ethernet network) and incorporates routing functionality to connect the control panel to the IT network.

\begin{figure}[!ht]
    \centering
    \includegraphics[draft=false,width=1.0\linewidth]{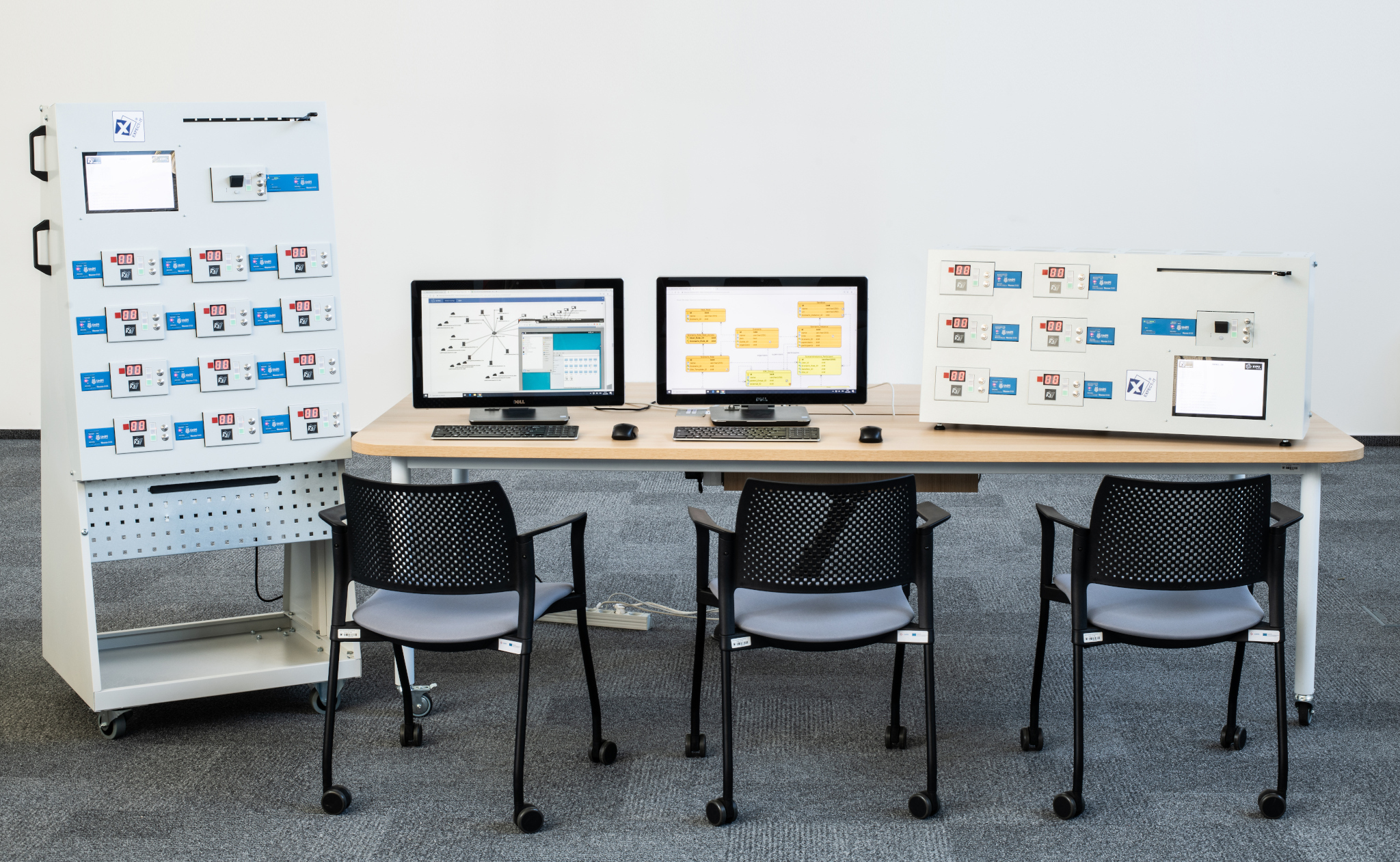}
    \caption{Physical hardware setup of the ICS testbed}
    \label{fig:ics-hw-setup}
\end{figure}

We built ten control panels to place at the top of the table and six as a movable trolley (see \Cref{fig:ics-hw-setup}). The tabletop setup is space-efficient, and the portable trolley provides mobility. The control panel is easy to handle; it requires only a power cord to connect to the mains electricity supply and Ethernet cable to connect to the IT network. The power consumption of one control panel is less than the power consumption of a desktop computer ($\leq$\,200\,W).

\subsection{Software Components}

\begin{figure}[!ht]
    \centering
    \includegraphics[width=0.9\linewidth]{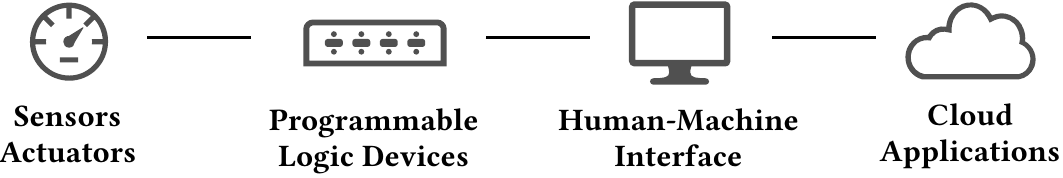}
    \caption{Interfaces between software components} 
    \label{fig:testbed-sw-architecture}
\end{figure}

The physical equipment provides fidelity of the operational environment, but software is needed to replicate the functions and behavior of various ICS systems. \Cref{fig:testbed-sw-architecture} shows the proposed software architecture based on the simplified ISA-95 model.

Based on our experience from developing and delivering hands-on cybersecurity courses, we defined the following requirements on software components of \KI testbed:


\begin{itemize}[leftmargin=15pt]
\item \emph{Open-source model} -- access to source code and full software control, no licensing fees and licensing obstacles, community support, and collaboration.
\item \emph{Operating system} -- a fully-fledged operating system with Raspberry PI support, operating-system-level virtualization, and high-speed networking.
\item \emph{Orchestration} -- the ability to manage all testbed devices - configuration management and application deployment, automated preparation of testbed environment.
\item \emph{Communication protocols} -- support for numerous legacy and emerging communication protocols used in ICS and IT environment.
\end{itemize}

The software stack of ICS testbed includes Linux OS (Debian optimized for PLC devices), Docker ecosystem~\cite{docker2019}, and on-premise OpenStack~\cite{openstack2019} cloud environment. We combine cloud deployment (virtual machines in OpenStack) with physical devices (PLCs, sensors, and actuators) to create ICS systems with varying levels of fidelity.

Automated orchestration of the testbed environment is of utmost importance. The central testbed controller runs as a virtual appliance. It provides management and monitoring of ICS testbed and contains Docker repository for PLC devices. The PLC devices are pre-installed with Debian OS and enabled Docker support. Using Docker containers simplifies software deployment and configuration of testbed components. 


The openness of the used software allows us to implement virtually any new software component. We focus on two use-cases: widely deployed systems and new emerging technologies. Communication protocols and application interfaces are essential to creating a complete ICS system. There are dozens of industrial protocols, and many new protocols are being proposed every year. Widely deployed protocols are Modbus and DNP3~\cite{Knapp2014book}. They have been used for decades for communications between ICS devices. The new emerging protocols represent MQTT~\cite{mqtt2019} and REST~\cite{Richardson2007rest}.


\section{ICS Cybersecurity Course Design} \label{sec:ics-course}

This section presents our proposed ICS cybersecurity course that employs the ICS testbed. While the previous section described the hardware and software components of the testbed, it did not deal with content. One of our motivations for this course, apart from student learning, is that the students will create training content for the testbed. When writing this section, we followed the guidelines for planning new courses~\cite{Walker:2016} and Joint Task Force on Cybersecurity Education (JTF) Cybersecurity Curricula 2017~\cite{cybered}.

\subsection{Course Goals and Covered Topics}

The overall goal of the course is to provide undergraduate students with an awareness of threats within the ICS domain via hands-on experience. As in the \textit{authentic learning} framework~\cite{lombardi2007authentic}, the focus is on solving real-world problems and learning by doing. The students' final product of the course is a training game for exercising both attacks at and defense of a selected industrial process. Our students previously created such games in the IT domain~\cite{kypolab-course}.

The primary JTF curriculum Knowledge Area (KA) the course covers is System Security, with Knowledge Units (KU) of Common System Architectures, System Thinking, and System Control. The secondary KAs are Component Security (KU Component Testing), Connection Security (KU Network Defense), Data Security (KU Secure Communication Protocols), and Organizational Security (KU Systems Administration). We also marginally include the KA Social Security (KUs Cybercrime and Cyber Law). Finally, the course focuses not only on technical skills but also enables students to exercise communication, presentation skills, and time management.


\subsection{Course Format}

The course is aimed at computer science university students, namely undergraduates with a basic background in computer networks and security. The recommended prerequisite is completing our Cyber Attack Simulation course in the IT domain~\cite{kypolab-course}. The initial run of the course is prepared for 6 students; however, the training facility described in \Cref{sec:testbed} can accommodate up to 20 students who can work in pairs using the 16 control panels (see \Cref{fig:ics-hw-setup}). The course spans the whole semester (13 weeks). It is taught in a flipped classroom format~\cite{bishop2013flipped} with 2-hour long weekly lab sessions, various homework assignments, and a hands-on semester project.

The necessary infrastructure includes, apart from the ICS testbed, also a CTF game infrastructure for running students' games (such as CTFd~\cite{chung2017} or KYPO cyber range platform~\cite{kypo-icsoft17}), and Gitlab repositories for students' projects. We appreciate the effort of the open-source community, such as learning resources, documentation, and countless projects~\cite{ics-awe,openplc,node-red}, which will help students to understand the used software. 

\begin{table*}
\caption{The schedule and the structure of the ICS cybersecurity course, along with the student deliverables and their contribution to the total course grade (with 10\% being for active participation in class), and important tasks for the instructor}
\label{table:course-structure}
\renewcommand\arraystretch{1.5}
\begin{tabular}{llll}
\hline
\textbf{Week} & \textbf{Class content} & \textbf{Student homework task (\% of the grade)} & \textbf{Instructor tasks} \\ \hline
 1 & Motivation, real attacks, legal issues & Prepare a presentation about an ICS attack (5\%) & --- \\ \hline
 2 & Student presentations of chosen attacks & Read this paper and some of the references & Grade the presentations \\ \hline
 3 & Hands-on labs on ICS testbed familiarization & Write an ICS security threat landscape report (5\%) & --- \\ \hline
 4 & Threat discussion, demo on ICS testbed & Write a short survey of CTF games in ICS (5\%) & Grade the reports \\ \hline
 5 & Merge surveys, introduce game concepts & Select threats for your game & Grade the surveys \\ \hline
 6 & Threat modeling, storyline, consultation & Write a game draft & Check the game drafts \\ \hline 
 7 & Preparing ICS part, educational objectives & Add learning outcomes and prerequisites & Check the game drafts \\ \hline
 8 & Preparing ICS and IT part & Prepare an alpha version of the game & Deploy the games \\ \hline
 9 & Dry run of the games with peers & Improve the game, submit bug reports (5\%) & Review bug reports \\ \hline
10 & Bug presentations, game improvement & Improve the game & --- \\ \hline
11 & Documentation, automation, deployment & Submit the game for presentation (50\%) & Deploy the games \\ \hline
12 & Public run of the games & Write a reflection from the public run (5\%) & Oversee the event \\ \hline
13 & Final reflections & Fix any issues that emerged in the public run (15\%) & Grade the games \\ \hline
\end{tabular}
\end{table*}

\subsection{Course Syllabus}

\Cref{table:course-structure} provides an overview of the course syllabus, student deliverables, and assessment methods. The course is divided into three parts: basics of ICS, development of an ICS training game, and its presentation and submission.

\subsubsection*{ICS Principles}

Since we expect the students to have little knowledge of ICS, the first class session will motivate the topic by presenting examples of past cyber attacks such as Stuxnet~\cite{langner2011}. The goal is to demonstrate the real-world impact of ICS incidents. We will follow by explaining the related terms, such as critical information infrastructures, and the corresponding legal regulations (such as a national Act on cybersecurity). For their homework, the students will individually choose a real, publicly-known attack on ICS and present it to others next class (in 15 minutes, including Q\&A). After the presentations, the homework assigned in week 2 will be reading this paper and the papers we reference in the related work.

In week 3, the students familiarize themselves with the ICS testbed. They will complete several hands-on labs to learn the basic operational features of HW and SW components of the testbed. At the end of the class, they will discuss in groups how to demonstrate the known attacks using ICS testbed. As an individual homework assignment, they will search for existing ICS security threat landscape reports/lists, like the OWASP Internet of Things Project~\cite{owasp-iot}.

The following week, each student will present their results. The group will discuss the severity of each threat, and which of them can or cannot be demonstrated on the \KI testbed to understand the capabilities and limitations of the testbed. The individual homework for the next week will be to prepare a 1-page written survey of CTF games in the ICS domain.

In the week 5 class, the students will engage in a pair activity of merging their reports to create a shared list of existing CTF games for the whole class. The motivation is to have a knowledge base of inspiration for students' games. The activity will follow with a short discussion centered around the question, ``What features should an engaging game have?'' The instructor will then briefly lecture on the principles of gamification~\cite{annetta2010} and provide an illustrative example to help students in their later assignment. The homework for the next week will be to think about a topic of student's game, which processes and threats the student will focus on, and how the student can use the ICS testbed for it. The instructor will highlight the specifics of ICS processes, and point out that they are threatened by different types of attacks than conventional IT systems. This homework starts the semester project phase.


\subsubsection*{Game Development}

Week 6 starts with an activity in which pairs of students ``peer-review'' each others' discovered threats using the Security Threat Modelling Cards~\cite{threatcards}. Students who finish will proceed to one-on-one consultations with the instructor to discuss the topic and the process of the game (output of the previous homework). Afterward, the students start working on the game narrative (storyline) and design the game flow, including the separation of tasks into levels. For their homework, the students will finish this design and send the draft to the instructor to receive formative feedback. The instructor will review the drafts and send comments before the next class.


In week 7, students will individually continue to develop their game, particularly the PLC-related part (Layer 1 of the ISA-95 architecture, see \Cref{fig:isa-95-architecture}). The instructor will then briefly lecture on the importance of the proper setting of learning outcomes and prerequisites, including examples from existing games. The students will use these instructions in their homework and add the learning outcomes and prerequisites to the description of their game.


Week 8 is dedicated to finishing the development of the PLC-related part and development and configuration of the Supervisory part (Layer 2 of the ISA-95 architecture). Students have to deliver an alpha version of their game for the dry run before the next class.

Week 9 starts with the dry run of students' games in pairs. Each student plays the game of another student for 45 minutes and takes notes about the learning experience. Then they switch roles. Afterward, the students are instructed on how to file a good bug report and report their feedback on the game in Gitlab. The instructor will review the submitted bug reports before the next class. The optional homework is to improve the games based on the dry run.


Week 10 starts with a short presentation of demonstrative examples of filed bug reports chosen by the instructor. For the rest of the class, students improve the games based on the feedback from the dry run.


In week 11, the students document their game and automate its deployment in the ICS testbed. They must submit the final version of their game three days before the next class, the course finale.

\subsubsection*{Game Presentation and Submission}

In week 12, the students take part in organizing a Hacking Day -- a public event during which other students of the university can play the created games. This event has two goals: motivating the students to work on their projects and popularizing ICS cybersecurity. Our experience from hosting such an event in the IT domain is described in~\cite{kypolab-course}.

Finally, week 13 is dedicated to students' reflections and the Hacking Day wrap-up in a focus group discussion. If any issues emerged in their game during the Hacking Day, they must fix them.

\section{Conclusions}
\label{sec:conclusions}

We shared the design details of KYPO4INDUSTRY, a testbed for teaching ICS cybersecurity in a hands-on way. Moreover, we proposed a novel university course that employs the testbed. The students will practically learn about threats associated with the ICS domain, develop an educational cyber game, and exercise their soft skills during multiple public presentations. The acquired skills will be essential for the computer science undergraduates who will be responsible for cybersecurity operations of an entire organization in their future career. We suppose that more organizations will employ cyber-physical systems, and so understanding of ICS-specific features will constitute an advantage for the prospective graduates.

\subsection{Experience and Lessons Learned}

Although using simple microprocessor systems (e.g., development boards) in teaching is popular, these systems do not replicate complex ICSs. Cyber-physical systems are unique and change with the physical process they control. The proposed testbed provides ten tabletop control panels and six mobile installations. In total, students can work with 148 PLCs, which use the popular Raspberry Pi single-board computers. The individual components (PLCs, sensors, actuators) are available off the shelf; however, the challenge is to build a hardware setup that will replicate the ICS in a laboratory environment. Addressing this challenge involves multiple engineering professions and requires external collaboration.


\subsection{Future Work}

The presented testbed is modular; therefore, it can be gradually upgraded as new advances in the field will emerge in the future. We rely on open-source components that are supported by large communities of users and developers. Still, there is room for future work on the content of training scenarios and novel instruction methods in the ICS domain. Another interesting research idea is to develop methods for creating cyber games and compare whether they work the same in the IT and ICS domain.

\begin{acks}
This research was supported by the \grantsponsor{ERDF}{ERDF}{} project \textit{CyberSecurity, CyberCrime and Critical Information Infrastructures Center of Excellence} (No. \grantnum{ERDF}{CZ.02.1.01/0.0/0.0/16\_019/0000822}).
\end{acks}


\bibliographystyle{ACM-Reference-Format}
\balance
\bibliography{references}

\end{document}